\documentclass[10pt,aps,twocolumn,superscriptaddress,noshowpacs,nofootinbib,noshowkeys,floatfix]{revtex4}
\usepackage[dvips]{graphicx}
\usepackage[colorlinks=true,linktocpage=true,linkcolor=blue,citecolor=blue]{hyperref}
\usepackage[usenames,dvipsnames]{color}
\usepackage{amsmath, amssymb}
\usepackage{multirow}
\usepackage{longtable}
\usepackage{bm}
\usepackage{color}
\usepackage[normalem]{ulem}  

\renewcommand\sout{\bgroup \color{blue} \ULdepth=-.5ex \ULset}

\begin{document}

\preprint{}

\title{Fireball tomography from bottomonia elliptic flow in relativistic heavy-ion collisions}

\author{Partha Pratim Bhaduri}
\affiliation{Variable Energy Cyclotron Centre, HBNI, 1/AF Bidhan Nagar, Kolkata 700 064, India}
\author{Mubarak Alqahtani}
\affiliation{Department of Basic Sciences, College of Education, Imam Abdulrahman Bin Faisal University, Dammam 34212, Saudi Arabia}
\author{Nicolas Borghini}
\affiliation{Fakult\"at f\"ur Physik, Universit\"at Bielefeld, Postfach 100131, D-33501 Bielefeld, Germany}
\author{Amaresh Jaiswal}
\affiliation{School of Physical Sciences, National Institute of Science Education and Research, HBNI, Jatni-752050, India}
\author{Michael Strickland}
\affiliation{Department of Physics, Kent State University, Kent, OH 44242 United States}

\date{\today}

\begin{abstract}
We calculate the elliptic flow of bottomonia produced in Pb$\,+\,$Pb collisions at $\sqrt{s_{\rm NN}}=5.02$~TeV. We consider temperature-dependent decay widths for the anisotropic escape of various bottomonium states and observe that the transverse momentum dependence of bottomonia elliptic flow provides a tomographic information about the QGP fireball at different stages of its evolution. For the space-time evolution of the fireball, we employ simulation results from the 3+1D quasiparticle anisotropic hydrodynamic model. We find that our results for transverse momentum dependence of bottomonia elliptic flow are in reasonable agreement with experimental results from the ALICE and CMS collaborations.
\end{abstract}



\maketitle

%
\section{Introduction}
\label{s:intro}

Relativistic heavy-ion collisions create a strongly interacting hot and dense medium of deconfined quarks and gluons called the quark-gluon plasma (QGP). The thermodynamic and transport properties of QGP are governed by quantum chromodynamics (QCD). Therefore, experimental results from high-energy nucleus-nucleus collisions at the CERN Large Hadron Collider (LHC) and the BNL Relativistic Heavy Ion Collider (RHIC) are important to test the predictions of QCD. There are several experimental observables for studying the QGP, one of which is the study of the in-medium propagation of heavy quarks (charm and bottom quarks) and their quarkonium bound states.  Heavy quark observables provide very useful internal probes of the QGP~\cite{Andronic:2015wma}. In particular, heavy quarkonia created in these collisions are significantly affected by the medium, leading to distinctive features in their spectra and yields. Therefore, the study of heavy quarks and quarkonia is important to understanding QGP evolution \cite{Matsui:1986dk, Karsch:1987pv, Brambilla:2004wf, vanHees:2004gq, vanHees:2005wb, Rapp:2008tf, Kluberg:2009wc, Andronic:2015wma}.

The study of the suppression of heavy quarkonia, as a signature for the creation of an equilibrated QGP, was first proposed by Matsui and Satz based on the in-medium dissociation of quarkonium bound states due to Debye screening of the quark-antiquark ($q\bar{q}$) potential~\cite{Matsui:1986dk}. In this simple classical picture, the quarkonia bound states having the largest binding energy, i.e., the $J/\psi$ and the $\Upsilon$, possess the highest dissociation temperatures~\cite{Matsui:1986dk, Karsch:1987pv}. However, in recent years, it has been shown from first-principles finite-temperature QCD calculations that the in-medium $q\bar{q}$ potential also contains an imaginary part which is generally associated with the in-medium breakup rate of quarkonium states~\cite{Brambilla:2008cx, Brambilla:2011sg, Brambilla:2013dpa, Laine:2006ns, Dumitru:2007hy}. This in-medium break-up rate leads to large thermal widths for quarkonia states, resulting in their dissociation even at temperatures at which they would be expected to survive in the traditional scenario~\cite{Strickland:2011mw, Strickland:2011aa, Krouppa:2015yoa, Krouppa:2016jcl, Krouppa:2017jlg, Margotta:2011ta}. 

Another important aspect in the study of heavy quarkonia is the role of the QGP dynamics. In an expanding QGP, the temperature drop leads to a decrease of the Debye screening of the potential, thereby resulting in recombination of uncorrelated $q\bar{q}$ pairs into a stable bound state~\cite{Grandchamp:2003uw, Emerick:2011xu, Thews:2000rj, BraunMunzinger:2000px}. While this process is quite significant for charmonia yields~\cite{Zhao:2010nk} at RHIC and LHC, recombination seems to be marginal for bottomonia owing to the fact that $b\bar{b}$ pairs are produced much less abundantly in initial hard scatterings than charmonia~\cite{Du:2017qkv, Krouppa:2018lkt}. Moreover, since dissociation is a non-instantaneous process, the survival probability of bottomonium states is in-medium path length dependent and is naturally described within an ``escape mechanism'' scenario~\cite{Borghini:2010hy, He:2015hfa, Romatschke:2015dha, Jaiswal:2017dxp, Bhaduri:2018iwr}. For the spatially anisotropic QGP produced in a generic high energy heavy-ion collision, this path-length dependence leads to an anisotropic emission pattern of quarkonia, which in turn results in anisotropic momentum distribution and therefore flow-like signatures, as was first predicted for $J/\psi$~\cite{Wang:2002ck}. 

Since bottomonia are very massive, it is expected that they are produced very early in the initial hard scattering and their propagation is not affected significantly by the medium collectivity. Therefore, the anisotropic escape mechanism would constitute the major contribution to the observed momentum anisotropy, which is usually expressed in terms of Fourier harmonics $v_n$. Moreover, since they are produced early with a spatial distribution given by hard binary collisions, their propagation through the medium and escape provide a tomographic probe of the fireball. In this work, we calculate the elliptic flow of bottomonia produced in Pb$\,+\,$Pb collisions at $\sqrt{s_{\rm NN}}=5.02$~TeV at the LHC. We consider temperature-dependent decay widths for anisotropic escape of various bottomonium states and demonstrate that the transverse momentum dependence of bottomonia elliptic flow reflects properties of the fireball at different stages of its evolution. 

%
\section{Hydrodynamical model}
\label{s:model_fireball}

For the initial spatial distribution of the  bottomonium states, we assume that, due to their large masses, they are produced in initial hard scattering when the heavy nuclei collide. Therefore, we distribute the bottomonium production points in the transverse plane according to the number of binary collisions, $N_{\rm coll}(x,y)$, obtained from the optical Glauber model. Along the longitudinal direction $(z)$, we assume a flat distribution. For the initial transverse momentum ($p_T$) distribution of the bottomonia, we adopt a power law dependence obtained from numerical PYTHIA simulations for proton-proton ($p+p$) collisions, scaled by the corresponding mass number of colliding nuclei.\footnote{We do not try to account for the difference between the parton distribution functions in a Pb nucleus and a proton. The effect of including nuclear PDFs (and their uncertainties) is expected to be small for $v_2$, while the reasonable agreement of our calculated $R_{AA}$ with the experimental results validate our approach~\cite{Vogt:2016eqd,Ferreiro:2018wbd}.}
The initial $\Upsilon$ distribution in $p+p$ collision from PYTHIA is given by \cite{Zhou:2014hwa}
\begin{equation}
\label{ppjpsi}
\frac{{\rm d}^2\sigma^{pp}_\Upsilon}{p_T\,{\rm d}p_T\,{\rm d}Y} =
\frac{4}{3\langle p_T^2\rangle_{pp}}\!\left(1+{\frac{p_T^2}{\langle p_T^2\rangle_{pp}}}\right)^{\!\!-3}{\frac{{\rm d}\sigma^{pp}_\Upsilon}{{\rm d}Y}} \, ,
\end{equation}
where $Y$ is the longitudinal rapidity in momentum space, while $\langle p_T^2\rangle_{pp}(Y)=20(1-Y^2/Y_{\max}^2)\;$(GeV/$c)^2$, with $Y_{\max}={\rm cosh}^{-1}(\sqrt{s_{\rm NN}}/(2m_{\Upsilon}))$ the maximum value of forward rapidity of the bottomonia. Finally, the momentum rapidity follows a Gaussian distribution given by:
\begin{equation}
\label{gaussian}
\frac{{\rm d}\sigma^{pp}_\Upsilon}{{\rm d}Y} = 
  \frac{{\rm d}\sigma^{pp}_\Upsilon}{{\rm d}Y}\bigg|_{Y=0} {\rm e}^{-Y^2/0.33Y_{\max}^2}.
\end{equation} 
In the present calculations, we integrate over $Y$ and consider the resulting $p_T$ distribution.

Owing to the quantum mechanical uncertainty principle, each bottomonium bound state has a finite formation time $\tau_{\rm form}$. In the bottomonium rest frame, the value of intrinsic formation time $\tau^0_{\rm form}$ is assumed to be inversely proportional to the binding energy in vacuum of each state. We use $\tau^0_{\rm form} = 0.2,~0.4,~0.6,~0.4,~0.6$ fm/$c$ for $\Upsilon(1S),~\Upsilon(2S),~\Upsilon(3S),~\chi_{b}(1P)$ and $\chi_{b}(2P)$ states, respectively~\cite{Krouppa:2016jcl}. Due to time dilation, the formation time in the plasma frame is given by $\tau_{\rm form} = E_T\tau^0_{\rm form} / M$, where $M$ is the mass of the bottomonium states and $E_T = \sqrt{p_T^2+M^2}$ is the transverse energy with $p_T$ being the transverse momentum. After their formation, the color-neutral $b\bar{b}$ bound states suffer few elastic scatterings in QGP, and due to their large mass, they propagate quasi freely with nearly straight-line trajectories. 

In order to accurately evaluate the survival of the bottomonium states propagating through the QGP medium, we need a realistic simulation of the space-time evolution of the expanding QGP. In the present work, we employ the simulation results from the recently developed 3+1D quasiparticle anisotropic hydrodynamics (aHydroQP) framework \cite{Alqahtani:2015qja}. This framework, which includes both shear and bulk viscosities as well as an infinite number of transport coefficients, has been quite successful in describing multitude of experimental observables for both Pb+Pb collisions at $\sqrt{s_{\rm NN}}=2.76$ TeV \cite{Alqahtani:2017jwl, Alqahtani:2017tnq} and Au+Au collisions at $\sqrt{s_{\rm NN}}=200$ GeV collisions \cite{Almaalol:2018gjh}. In this version of the aHydroQP code, the shear viscosity to entropy density ratio ($\eta/s$) is assumed to be constant and all other transport coefficients are determined self-consistently within a quasiparticle frameowrk. The temperature dependence of the quasiparticle mass is extracted from lattice QCD results for the entropy density \cite{Borsanyi:2010cj}. For details of this framework we refer the reader to Ref.~\cite{Alqahtani:2017mhy} which presents a comprehensive review of the approach.

The dissipative hydrodynamical evolution of QGP is started at $\tau_i = 0.25$ fm/c. In order to construct the initial energy density profile in the transverse plane, we use optical Glauber model  with a linear combination of binary collisions and wounded-nucleon. The fraction of binary collisions is set to $\kappa_{\rm binary} = 0.15$. The nucleon-nucleon inelastic collision cross section is taken to be $\sigma_{\rm NN} = 67$ mb. For the density distribution of the incoming nuclei, we use a standard Woods--Saxon profile, with saturation density $n_{0}=0.17$ fm$^{-3}$, nuclear radius $R_A = (1.12A^{1/3} - 0.86A^{-1/3})$ fm and skin depth $d = 0.54$ fm. We take the initial central temperature to be $T_0 = 630$ MeV and the specific shear viscosity to be $4 \pi\eta/s = 2.0$.  These values were determined by fits to identified particle spectra and successfully reproduce many bulk observables in 5.02 TeV Pb-Pb collisions ~\cite{Alqahtani:2020paa,Acharya:2019yoi}. The space-time evolution of the effective temperature in each centrality class was obtained by using 4D interpolating functions constructed from the aHydroQP model evolution for a given centrality. For details concerning the hydrodynamical framework and fits to 2.76 TeV experimental data, please see Ref.~\cite{Alqahtani:2020paa}.

\begin{figure}[t!]
\begin{center}
\includegraphics*[width=\linewidth]{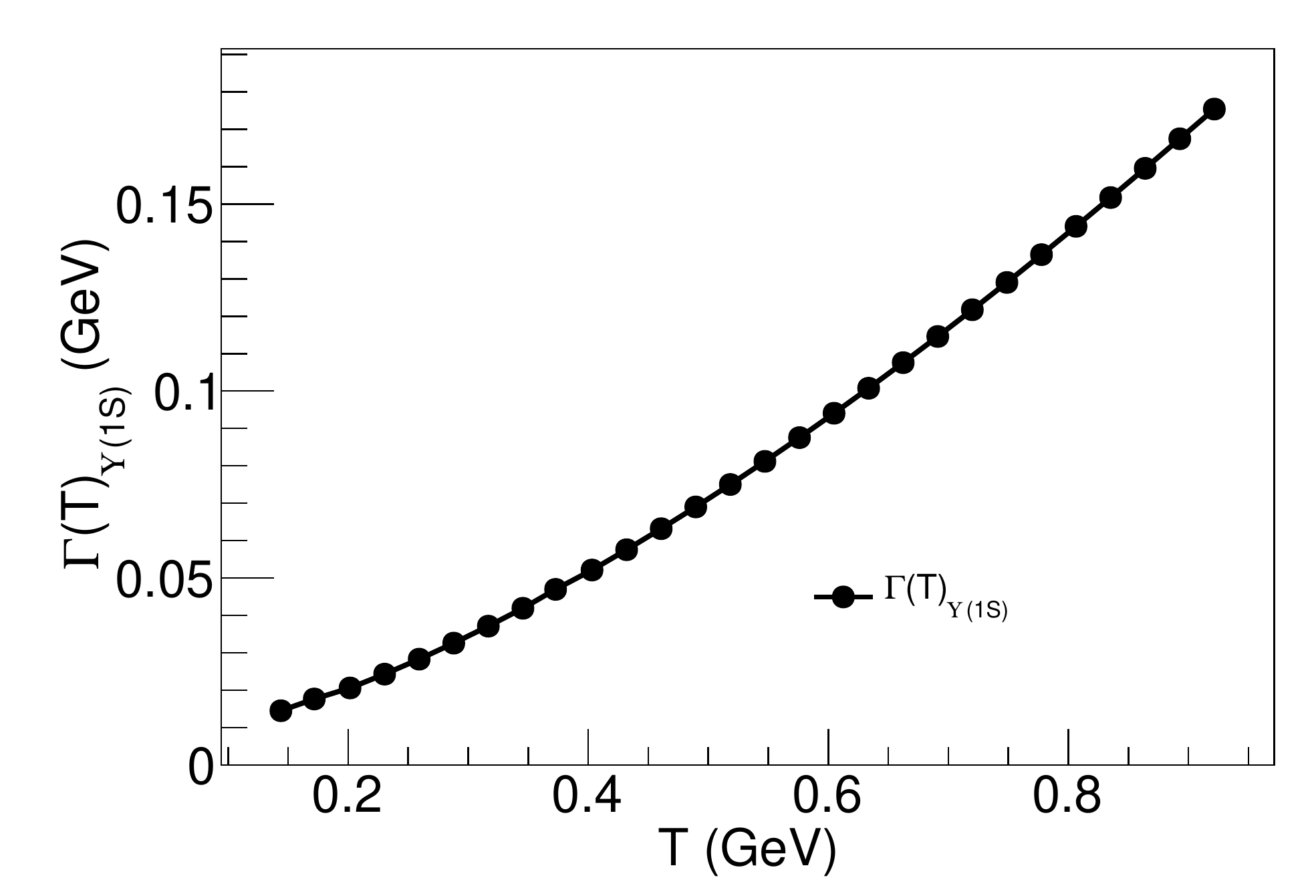}
\end{center}
\vspace{-0.5cm}
\caption{Temperature dependence of the in-medium decay width ($\Gamma(T)$) of $\Upsilon(1S)$ states.} 
\label{gamma_y1s}
\end{figure}

Once we know the temperature of the medium along the bottomonium trajectory from the aHydroQP model, one can calculate the breakup rates of the bottomonium states by using temperature-dependent thermal decay widths. We adopt the in-medium dissociation of different bound $b\bar{b}$ states from recent state-of-the-art numerical solution of 3D Schr\"odinger equation with a temperature-dependent complex heavy-quark potential \cite{Strickland:2011aa, Margotta:2011ta}. This potential is based on a generalization of Karsch--Mehr--Satz potential \cite{Karsch:1987pv} obtained from the internal energy. It also includes an imaginary part emerging from Landau damping of the exchanged gluons within the framework of hard-thermal-loop approximation~\cite{Laine:2006ns, Dumitru:2007hy, Strickland:2011aa}. This imaginary part of the potential results in a temperature-dependent in-medium breakup disassociation rate for each state.  The change in sign of the real part of the binding energy of a given state, on the other hand, allows one to determine when the state becomes completely unbound. For completeness, we illustrate in Fig.~\ref{gamma_y1s} the temperature dependence of the decay width ($\Gamma(T)$) of directly produced $\Upsilon(1S)$ states, as used in our present calculations.  

Note that, we set the temperature scale to $T_c = 160$ MeV, below which we assume that the screening effects due to plasma rapidly decreases because of transition to the hadronic phase. This temperature corresponds to the temperature of the assumed QGP phase transition. This is a reasonable approach since, in the hadronic phase, the widths of the bottomonium states become approximately equal to their vacuum widths, which are negligible in comparison to the in-medium widths. Taking all these aspects into consideration, the thermal decay width of formed bottomonia is modeled as 
\begin{equation}\label{Gamma}
\Gamma(T(x,y,\tau)) = 
\left\{
\begin{array}{ll}
2 \Im  & \qquad {\rm for} \quad \Re >0, \\
10\;{\rm GeV}  & \qquad {\rm for} \quad \Re \le 0, \\
\end{array}
\right.
\end{equation}
where we use the notations $\Re\equiv{\rm Re}[E_{\rm bind}(T(x,y,\tau))]$ and $\Im\equiv{\rm Im}[E_{\rm bind}(T(x,y,\tau))]$ for the real and imaginary parts of the in-medium binding energy, respectively. The temperature depends on the instantaneous transverse position $(x,y)$ of the bottomonium state at time $\tau$ along its trajectory. The value of 10 GeV in the second case in Eq.~\eqref{Gamma} is chosen to ensure that there is rapid dissociation of states which are fully unbound. However, in practice, the results are insensitive to this value provided that it is large enough to quickly dissociate the state under consideration.

In order to account for the finite time for bound states to develop from the primordial $b\bar{b}$ wave packet, we include a formation-time effect. This effect tends to decrease the suppression rate which one can intuitively associate with a geometric expansion of the $b\bar{b}$ wave packet from almost point-like production to the bound-state size~\cite{Du:2017qkv}. Therefore we correct the decay width accordingly as 
\begin{equation}\label{Gamma_eff}
\Gamma^{\rm eff}(T) = 
\left\{
\begin{array}{ll}
\Gamma(T)\,\tau/\tau_{\rm form}  & \quad {\rm for}~ \tau < \tau_{\rm form}, \\
\Gamma(T)  & \quad {\rm for}~ \tau \geq \tau_{\rm form}, 
\end{array}
\right.
\end{equation}
where, $\tau_{\rm form}$ is the bottomonium formation time in the plasma rest frame. We further assume that bottomonia only dissociate when the QGP is formed, i.e.\ $\Gamma^{\rm eff}(T) = 0$ for $\tau<\tau_i$. 

In order to study the propagation of bottomonium states, we consider that the QGP is formed at $\tau_i$ and the bottomonia are created with transverse coordinates $(x,y)$ and transverse momentum $p_T$ along the azimuthal $\phi_p$ direction. The position of this state at any future time $\tau$ is then given by\footnote{For a given $(x,y)$ in the transverse plane and time $\tau$, the temperature varies only weakly in the rapidity direction due to the approximate boost invariance of the QGP created at high energies.  For this reason, we do not indicate the $z$ direction.}
\begin{equation}\label{future_position}
x' = x + v_T\tau'\cos\phi_p\,, \qquad y' = y + v_T\tau'\sin\phi_p\,,
\end{equation}
where $v_T\equiv p_T/E_T$ is the bottomonium transverse velocity and $\tau'\equiv\tau - \tau_i$. The instantaneous temperature along a given bottomonium trajectory are obtained from the aHydroQP model predictions for the full 3+1D evolution of the QGP effective temperature. Subsequently, the effective in-medium decay width is determined for these temperatures using Eqs.~\eqref{Gamma}~and~\eqref{Gamma_eff}.

The in-medium decay width, as obtained above, determines the survival probability of a given state as it propagates through the medium. The final transmittance probability for a bottomonium state, labeled by $j$, is given by~\cite{Strickland:2011mw, Strickland:2011aa, Krouppa:2015yoa, Krouppa:2016jcl}
\begin{equation}\label{transmittance_final}
{\cal T}_j(x,y,p_T,\phi_p) = \exp\!\left[-\!\int_{\tau_i}^{\tau_f}\! {\rm d}\tau'\;\Gamma^{\rm eff}_j \bigg(T(x', y';\,\tau')\bigg)\right],
\end{equation}
where the final time $\tau_f$ is determined in the aHydroQP model as the proper time at which the local temperature of the expanding medium drops below the freeze-out temperature $T_f = 130$ MeV. Using the above equation, we calculate the transmitted spectra as a function of the azimuthal angle and the transverse momentum of all of the produced bottomonium states,
\begin{equation}\label{spectra_pt_phi}
\dfrac{{\rm d}N_j}{p_T\,{\rm d}p_T\,{\rm d}\phi_p}=\!\int\! {\rm d}x\,{\rm d}y\,N_{coll}(x,y)\frac{{\rm d}^2\sigma^{pp}_\Upsilon}{{\rm d}^2p_T\,}{\cal T}_j(x,y,p_T,\phi_p) \, .
\end{equation}
Using the above equation, one can calculate the nuclear modification factor and elliptic flow of the $j$-th resonance state as a function of transverse momentum using the standard definitions.

\begin{figure}[t!]
\begin{center}
\includegraphics*[width=\linewidth]{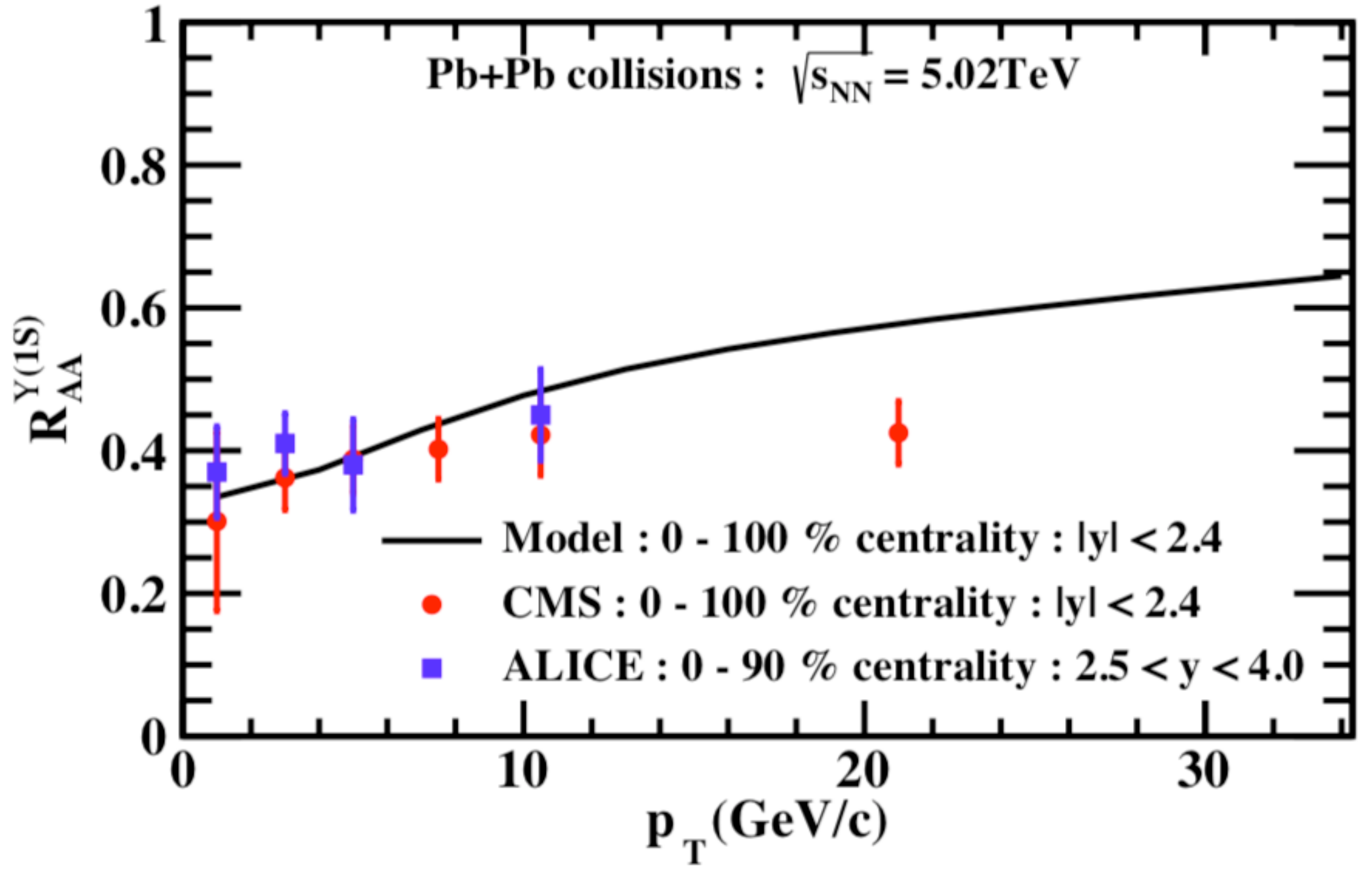}
\end{center}
\vspace{-0.5cm}
\caption{Transverse momentum dependence of nuclear modification factor $R_{AA}(p_{T})$ for $\Upsilon (1S)$ state for Pb$\,+\,$Pb collisions at $\sqrt{s_{\rm NN}}=5.02$~TeV in 0--90\% centrality class for ALICE and 0--100\% centrality class for model calculation and CMS. The model calculations and the CMS measurements are at mid-rapidity and the ALICE measurements are at forward rapidity ($2.5 < y< 4.0$). For ALICE and CMS data points the statistical and systematic errors are added in quadrature.} 
\label{RAA_comp}
\end{figure}

%
\section{Results and discussions}
\label{s:results}


In order to calculate the nuclear modification factor $R_{\rm AA}$ and elliptic flow $v_2$ of bottomonia due to escape probability through an anisotropic medium, we numerically evaluate Eq.~\eqref{spectra_pt_phi}. We consider Pb$\,+\,$Pb collisions at $\sqrt{s_{\rm NN}}=5.02$~TeV in various centrality classes. We obtain the ``raw'' spectra for each bottomonium state by integrating over the transverse temperature profile in Eq.~\eqref{spectra_pt_phi}. The feed down contribution from excited states are calculated using a $p_{T}$-averaged feed down fraction taken from a recent compilation of $p+p$ data at LHC \cite{Krouppa:2016jcl}: $f_{2S\rightarrow 1S}=8.6\%$, $f_{3S\rightarrow 1S}=1\%$, $f_{1P\rightarrow 1S}=17\%$, $f_{2P \rightarrow 1S}=5.1\%$ and $f_{3P \rightarrow 1S}=1.5\%$. Subsequently, the inclusive spectrum for $\Upsilon (1S)$ is obtained by linear superposition of the raw spectra for each state and given by
\begin{equation}\label{in_spec}
\frac{{\rm d}N^{\rm all}_{\Upsilon(1S)}}{p_T\,{\rm d}p_T\,{\rm d}\phi_p} = \sum_j f_j\frac{{\rm d}N_j}{p_T\,{\rm d}p_T\,{\rm d}\phi_p}.
\end{equation}
The inclusive spectrum thus obtained is then used to calculate the nuclear modification factor and elliptic flow of $\Upsilon(1S)$. 

\begin{figure}[t!]
\begin{center}
\includegraphics*[width=\linewidth]{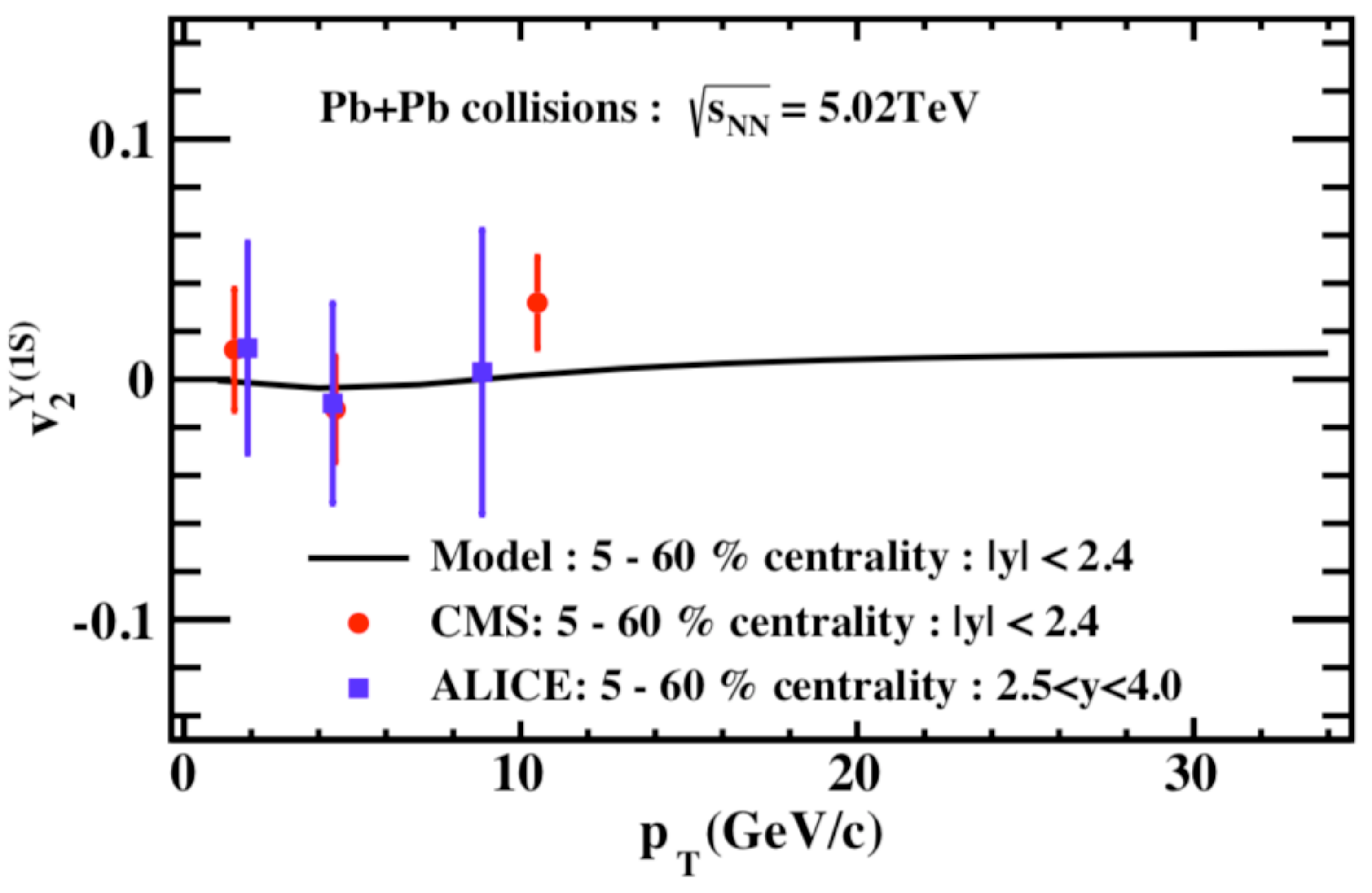}
\end{center}
\vspace{-0.5cm}
\caption{Transverse momentum dependence of elliptic flow of the $\Upsilon (1S)$ state in Pb$\,+\,$Pb collisions at $\sqrt{s_{\rm NN}}=5.02$~TeV in 5--60\% centrality class. The model calculations and the CMS data points are at mid-rapidity ($|y| < 2.4$) and the ALICE data points are at forward rapidity ($2.5 < y< 4.0$). For the ALICE and CMS data points the statistical and systematic errors have been added in quadrature.} 
\label{v2_comp}
\end{figure}

In Fig.~\ref{RAA_comp}, we show our model predictions for the transverse momentum dependence of nuclear modification factor $R_{AA}(p_{T})$ of $\Upsilon (1S)$ in Pb$\,+\,$Pb collisions at $\sqrt{s_{\rm NN}}=5.02$~TeV, together with results from ALICE (0--90\% centrality class, forward rapidity $2.5 < |y| < 4$)~\cite{Acharya:2018mni} and CMS (0--100\% centrality class, mid-rapidity $|y| < 2.4$)~\cite{Sirunyan:2018nsz}. The model calculations are performed in the same $|y| < 2.4$ rapidity interval as CMS in ten centrality bins: 0--10\%, 10--20\%, $\cdots$, 90--100\% centrality. The results are averaged with a weight proportional to $e^{-C/20}$ ($C$ being the centre point of the respective centrality bin) and accounting for the inclusive $\Upsilon (1S)$ yield in each bin, yielding our reported results for the 0--100\% centrality interval \cite{Krouppa:2016jcl}. Our calculated $R_{AA}$ is in reasonable agreement with the available measurements from both CMS and ALICE.  Although our calculations were performed at central rapidity, the weak rapidity dependence of charged particle multiplicity results in similar dynamics at mid- and forward rapidities~\cite{Adams:2017}.  We note that, comparing to the previous calculations of $R_{AA}$ of $\Upsilon(1S)$ within the anisotropic hydrodynamics framework~\cite{Krouppa:2016jcl}, these prior results did not include pre-resonance absorption. The effect of the pre-resonance absorption is larger for excited states, which have a longer formation time, and at higher $p_T$, due to time dilation. 

\begin{figure}[b!]
\begin{center}
\includegraphics[width=0.8\linewidth]{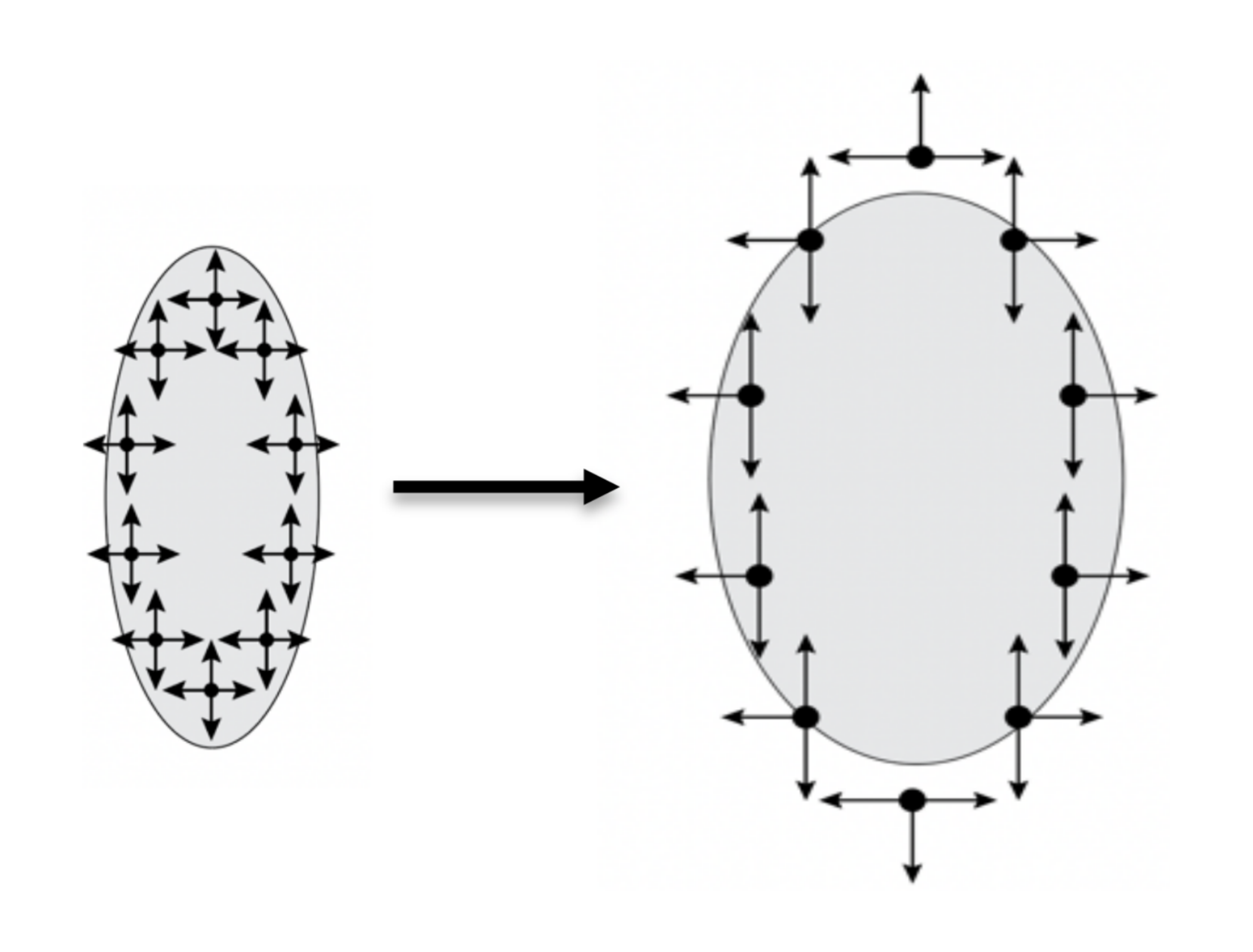}
\end{center}
\vspace{-0.5cm}
\caption{\label{cartoon}Schematic representation of buildup of negative elliptic flow at low $p_T$. The anisotropic expansion of the medium along $x$-axis leads to more suppression of low $p_T$ bottomonium states along the $x$-axis compared to $y$-axis due to states being overtaken by the expanding QGP.} 
\end{figure}

In Fig.~\ref{v2_comp}, we show our results for the transverse momentum dependence of elliptic flow parameter for $\Upsilon (1S)$ state, together with results from  ALICE~\cite{Acharya:2019hlv} and CMS~\cite{Sirunyan:2020qec}. Both model calculations and CMS results are for the 5--60\% centrality class and at mid-rapidity ($|y| < 2.4$), while the ALICE measurements are for the same centrality interval but at forward rapidity ($2.5 < y< 4.0$). Our calculations predict a negative $v_2$ for low $p_T$, which could be explained as follows. The bulk of bottomonia which survive the QGP are those produced near the ``edge'' of the fireball, see Fig.~\ref{cartoon}. This stems from the fact that the binary overlap profile produces pairs with elliptic initial geometry depending on the impact parameter. Those with momentum pointing into the QGP are strongly suppressed during their traversal. In the transverse expansion, the eccentricity of the fireball tends to decrease due to faster expansion along $x$-axis and can even overtake states which were previously ``escaped''. As a result, the low-momentum bottomonia undergo more suppression along the $x$-axis as they have to travel larger effective distance in the medium.\footnote{We note that, although we agree with the magnitude of $\Upsilon(1s)$ $v_2$ predicted by other models \cite{Du:2017qkv,Yao:2020xzw,Yao:2018sgn,Hong:2019ade,Ke:2018tsh}, our model has a clear qualitative difference in that it predicts a negative $v_2$ at low $p_T$.}

\begin{figure}[b!]
\begin{center}
  \includegraphics*[width=\linewidth]{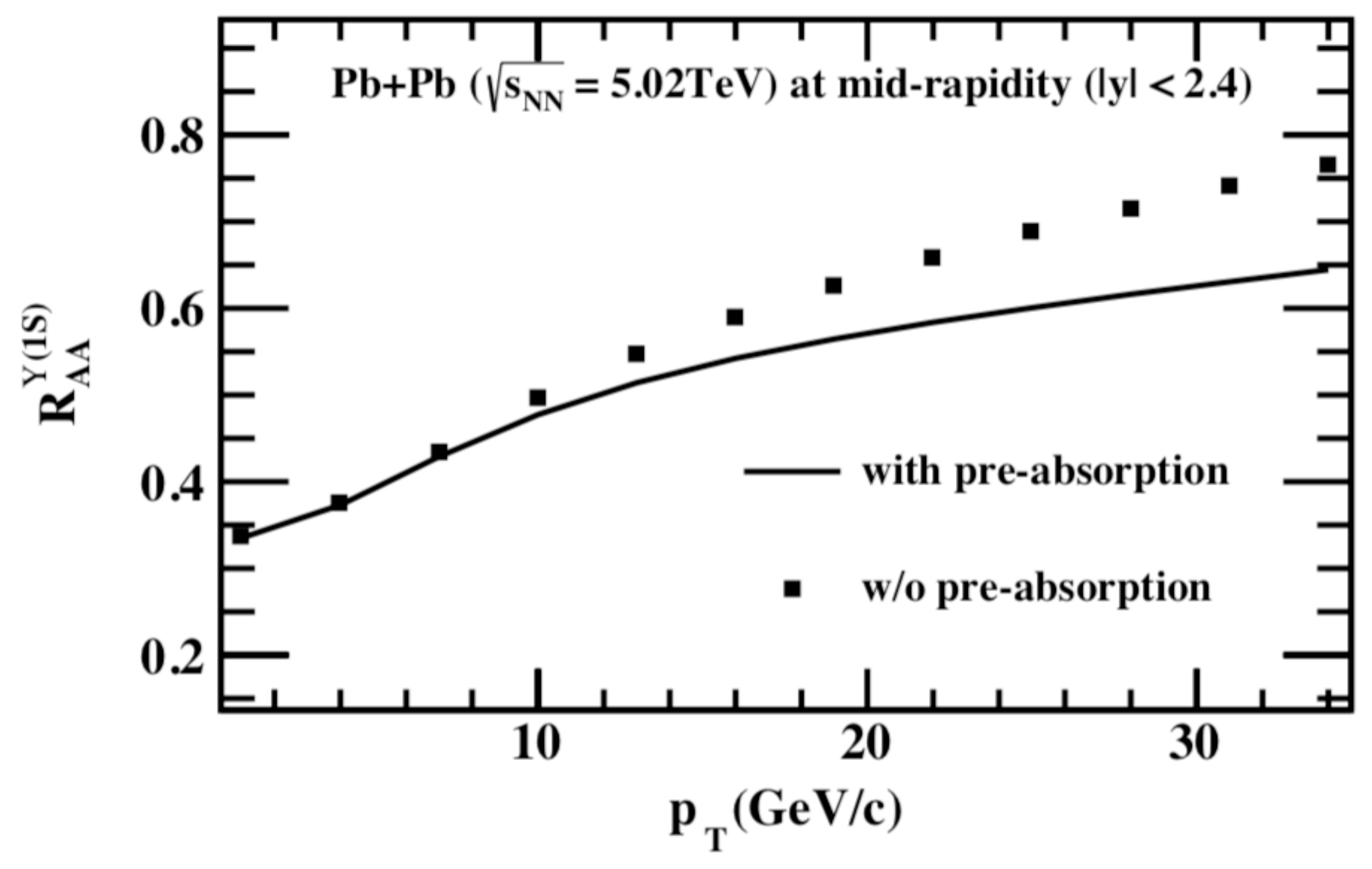}
  \includegraphics*[width=\linewidth]{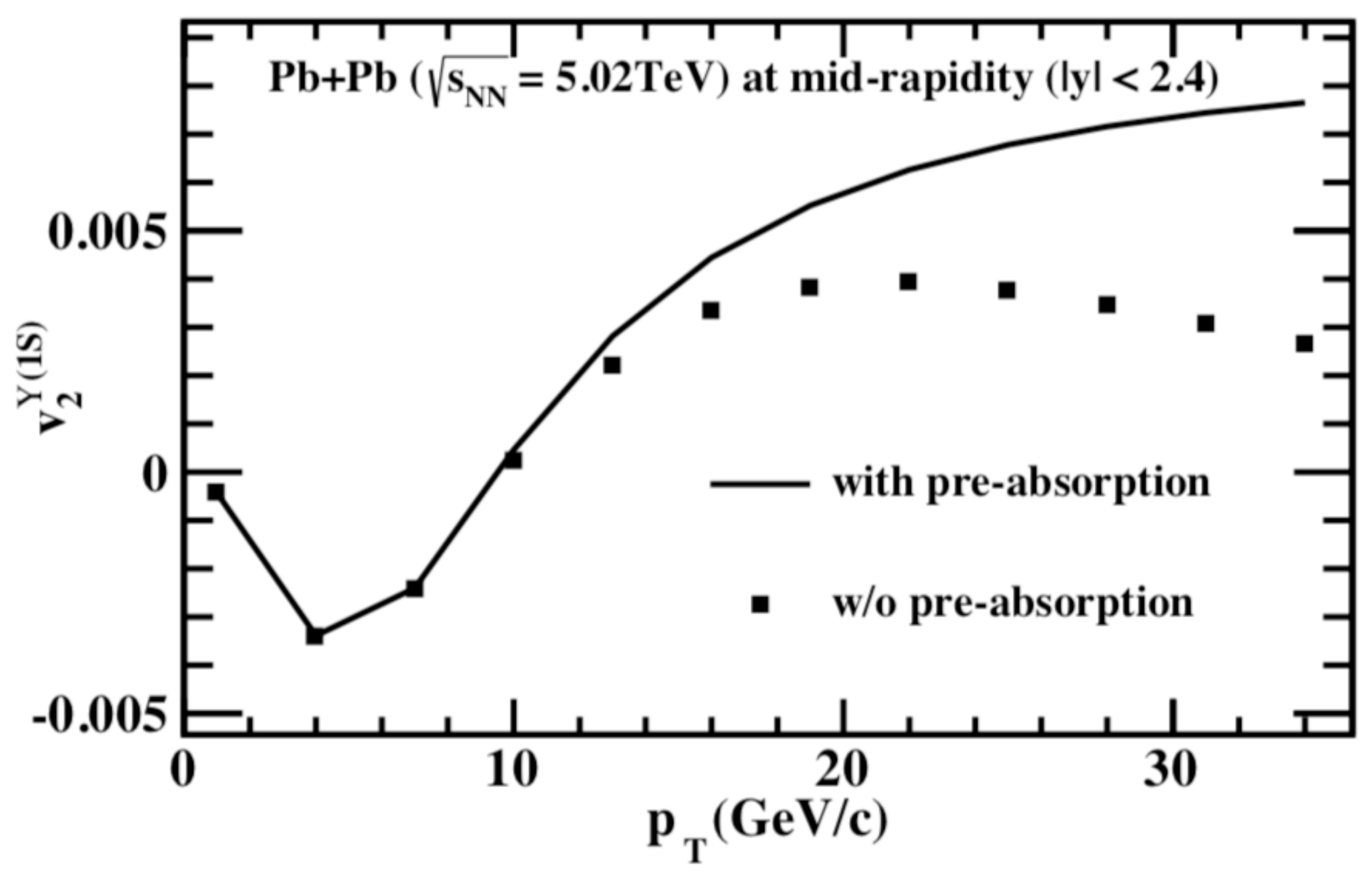}
\end{center}
\vspace{-0.5cm}
\caption{\label{preabs} The impact of pre-resonance absorption on the $p_{T}$ dependence of $R_{AA}$ (top) and $v_{2}$ (bottom) of the inclusive $\Upsilon (1S)$ states.} 
\end{figure}

On the other hand, bottomonia with high transverse momentum quickly traverse the fireball and feel the initial, predominantly elliptic, geometry. This leads to positive elliptic flow at large $p_T$ as shown in Fig.~\ref{v2_comp}. In order to explicitly verify our argument of negative $v_2$ at low $p_T$, in presence of transverse expansion, we calculated the $v_2$ of directly produced $\Upsilon(1S)$ states by considering only longitudinal expansion of the fireball. In absence of transverse expansion the negative component of $v_2$ at low $p_T$ vanishes and $v_2 \rightarrow 0$ as $p_T \rightarrow 0$, a feature common for all bottomonium states. Therefore, the transverse-momentum dependence of elliptic flow can provide valuable information about the time evolution of the fireball's shape in the transverse direction. This is analogous to taking snapshots of fireball anisotropy at different proper time which is sampled differently by states with different $p_T$. With this interpretation, one sees that bottomonium elliptic flow can provide tomographic information about the evolution of the transverse geometry of the fireball.


Before we conclude, let us briefly discuss the sensitivity of the predicted observables to systematic uncertainties associated with our model. The intrinsic formation times of the different bottomonium states are assumed to be inversely proportional to their vacuum binding energies. But the choice is not unique and highly model dependent. To quantify the resulting uncertainties on $R_{AA}(p_{T})$ and $v_{2}(p_{T})$ of the inclusive $\Upsilon (1S)$ states, we vary the intrinsic formation time of each resonance state by $\pm 50 \%$, from their default values. We find that the resulting change of either observable is small. Generically, decreasing the formation time results in a larger suppression, i.e. a smaller $R_{AA}$, and a slightly higher $v_2$. This affects bottomonia with a large $p_T$, as already observed in our study at $\sqrt{s_{\rm NN}} = 2.76$ TeV~\cite{Bhaduri:2018iwr}.
Accounting for bottomonia from the recombination of independent $b$ and $\bar{b}$ will also have a very small impact, since the mechanism is responsible for only a small fraction of the total bottomonium yield. 
 The variation of initial conditions like $\tau_i$, spatial profile, viscosity etc. of the hydrodynamic evolution are beyond the scope of the present work. As mentioned earlier, these parameters are adopted from~\cite{Alqahtani:2020paa,Acharya:2019yoi}, where they are determined by fitting the identified hadron spectra and can explain a multitude of bulk observables in 5.02 TeV Pb+Pb collisions. The relative motion between the bottomonia and the expanding QGP has not taken into consideration while calculating the in-medium decay widths. As discussed in~\cite{Bhaduri:2018iwr}, the effect on the estimated $v_{2}$ of the correction to the dissociation rates due to Doppler shifting~\cite{Escobedo:2013tca,Hoelck:2016tqf} is expected to be small, especially at small $p_T$. Finally we examine the role of pre-resonance absorption, that remains operative during the spatial expansion of the initially compact $b\bar{b}$ pair to a full grown resonance state. The choice of $\Gamma(T)^{\rm eff}$ is motivated from the linear increase in time of the size of the $Q\bar{Q}$ dipole. To test the influence of the ansatz, we show in Fig.~\ref{preabs} how turning off the dissociation of the pre-resonant $b\bar{b}$ pairs affects the shapes of $R_{AA}$ and $v_2$ of the inclusive $\Upsilon (1S)$ states. The suppression of the nascent pairs becomes important for the ``fast'' ($p_T > m$) bottomonium states, which in the absence of pre-resonance absorption can escape the plasma before they are suppressed.

\section{Summary and conclusion}
\label{s:summary}
%
%
In this paper, we calculated the elliptic flow of bottomonia produced in Pb$\,+\,$Pb collisions at $\sqrt{s_{\rm NN}}=5.02$~TeV at the Large Hadron Collider. We employed PYTHIA and the Glauber model to generate initial distribution of momentum and position of produced bottomonia. For the space-time evolution of the fireball, we employed a recently developed 3+1D quasiparticle anisotropic hydrodynamic model which has been tuned to reproduce experimental observables such has identified hadron spectra and elliptic flow at $\sqrt{s_{\rm NN}}=5.02$~TeV. We considered the effect of temperature-dependent decay widths on the traversal of various bottomonium states through a hot QGP medium with an spatially anisotropic geometry. We have also accounted for the feed down contribution from higher excited states to the bottomonium ground state. We propose that the transverse momentum dependence of bottomonia elliptic flow provides tomographic probe of the QGP fireball transverse expansion. We found that our results for transverse-momentum dependence of bottomonia elliptic flow are in good agreement with experimental results from the ALICE and CMS collaborations.
They also agree with recent calculations~\cite{Islam:2020gdv,Islam:2020bnp} simulating the real time quantum evolution of bottomonium states in the QGP, relaxing the assumption of an adiabatic evolution of the $b\bar{b}$ system~\cite{Dutta:2012nw}.
Looking forward, one can anticipate greatly increased statistics in future runs from the experimental collaborations at LHC.  Hopefully, with these increased statistics, we can more quantitatively assess the efficacy of our model.


\medskip

\begin{acknowledgments}
%
The authors would like to thank Jaebeom Park, Yongsun Kim and Vineet Kumar for useful discussions on CMS data. N.\,B.\ acknowledges support by the Deutsche Forschungsgemeinschaft (DFG, German Research Foundation) through the CRC-TR 211 'Strong-interaction matter under extreme conditions' -- project number 315477589 -- TRR 211. A.\,J.\  is supported in part by the DST-INSPIRE faculty award under Grant No. DST/INSPIRE/04/2017/000038. M.\,A.\  is supported by the Deanship of Scientific Research at the Imam Abdulrahman Bin Faisal University under grant number 2020-080-CED. M.\,S. is supported by the U.S. Department of Energy, Office of Science, Office of Nuclear Physics under Award No.~DE-SC0013470. This research was supported by the Munich Institute for Astro- and Particle Physics (MIAPP) of the DFG cluster of excellence ``Origin and Structure of the Universe''.  P.P.B would like to thank grid computing laboratory, VECC, Kolkata for facilitating the computation related to this work. 
\end{acknowledgments}


\end{document}